\documentstyle[epsf]{elsart}
\newcommand{\mafigura}[4]{
  \begin{figure}[hbtp]
    \begin{center}
      \epsfxsize=#1 \leavevmode \epsffile{#2}
    \end{center}
    \caption{#3}
    \label{#4}
  \end{figure} }
\newcommand{\eq}[1]{eq.~(\ref{#1})}
\newcommand{\as}{\alpha_s}

\newcommand{\beq}{\begin{equation}}
\newcommand{\eeq}{\end{equation}}
\newcommand{\bea}{\begin{eqnarray}}
\newcommand{\eea}{\end{eqnarray}}
\newcommand{\non}{\nonumber}
\newcommand{\msbar}{$\overline{\mathrm{MS}}$ }
\begin{document}
\begin{frontmatter}

{
\flushright 
FTUV/97-80\\
IFIC/97-89\\
TTP97-26\\
}
\vspace{0.5cm}

\title{$\alpha_s(m_Z)$ from $\tau$ decays 
with matching conditions \protect\newline at three loops} 

\author[Karlsruhe]{Germ\'an Rodrigo\thanksref{th:Valencia}},
\author[Valencia]{Antonio Pich} and
\author[Valencia]{Arcadi Santamaria} 
\address[Karlsruhe]{Inst. f\"ur Theoretische Teilchenphysik,
Universit\"at Karlsruhe\\ D-76128 Karlsruhe, Germany}
\address[Valencia]{Departament de F\'{\i}sica Te\`orica, IFIC, 
CSIC-Universitat de Val\`encia,\\ E-46100 Burjassot, Val\'encia, Spain} 
\thanks[th:Valencia]{On leave from Departament de F\'{\i}sica Te\`orica,
Universitat de Val\`encia, Val\`encia, Spain}

\begin{abstract}
Using the recent four-loop calculations of the QCD $\beta$-function
and the three-loop matching coefficients we study the induced errors in
$\as(m_Z)$ obtained from $\as(m_{\tau})$ due to the evolution procedure.
We show that, when consistent matching and running is used at this order,
these errors are pushed below $0.0005$ in $\as(m_Z)$. 
\end{abstract}

\end{frontmatter}

The beta function and the quark mass anomalous dimension
govern the evolution of the strong coupling constant 
and the quark masses through the renormalization group (RG) equations,  
\bea
& & \frac{da}{d \log \mu^2} = \beta(a) = - a^2 \left( \beta_0 +
\beta_1 a + \beta_2 a^2 + \beta_3 a^3 \right) + O(a^6)~, \non \\
& & \frac{d \log \bar{m}_q}{d \log \mu^2} = \gamma_m(a) =  - a \left( \gamma_0 +
\gamma_1 a + \gamma_2 a^2 + \gamma_3 a^3 \right) + O(a^5)~,
\label{RGE}
\eea
where $a=\as/\pi$ and $\bar{m}_q$ is the running mass of the quark $q$.
The coefficients of the  
QCD beta function have been calculated recently in the
\msbar scheme up to four loops~\cite{ritbergen.vermaseren.ea:97}
\bea
\beta_0 &=& \frac{1}{4} \left[11-\frac{2}{3}n_f\right]~, \qquad
\beta_1 = \frac{1}{16} \left[102-\frac{38}{3}n_f\right]~, \non \\
\beta_2 &=& \frac{1}{64} \left[\frac{2857}{2}-\frac{5033}{18}n_f
+\frac{325}{54}n_f^2 \right]~, \non \\
\beta_3 &=& \frac{1}{256} \left[
\left( \frac{149753}{6}+3564 \zeta_3 \right)
-\left( \frac{1078361}{162}+\frac{6508}{27}\zeta_3 \right)n_f
\right. \non \\ & & \left.
+ \left( \frac{50065}{162}+\frac{6472}{81}\zeta_3 \right)n_f^2
+\frac{1093}{729}n_f^3 \right]~,
\eea
and also the coefficients of the quark-mass anomalous
dimension have been calculated
at the same order~\cite{chetyrkin:97,vermaseren.larin.ea:97},
\bea
\gamma_0 &=& 1~, \qquad
\gamma_1 = \frac{1}{16} \left[\frac{202}{3}-\frac{20}{9}n_f\right]~, \non \\
\gamma_2 &=& \frac{1}{64} \left[1249+
\left( -\frac{2216}{27}-\frac{160}{3} \zeta_3 \right)n_f
-\frac{140}{81}n_f^2 \right]~, \non \\
\gamma_3 &=& \frac{1}{256} \left[\frac{4603055}{162}
+\frac{135680}{27}\zeta_3 - 8800\zeta_5 \right. \non\\
&+& \left(-\frac{91723}{27}-\frac{34192}{9}\zeta_3 + 880\zeta_4
+\frac{18400}{9}\zeta_5 \right) n_f \non\\ 
&+&\left.\left( \frac{5242}{243}+\frac{800}{9}\zeta_3
-\frac{160}{3}\zeta_4\right) n_f^2
+\left( -\frac{332}{243}+\frac{64}{27}\zeta_3 \right) n_f^3 \right]~. 
\eea
Here $\zeta_n$ is the Riemann zeta-function (
$\zeta_2=\pi^2/6$, $\zeta_3 = 1.202056903 \ldots$,
$\zeta_4 = \pi^4/90$ and $\zeta_5 = 1.036927755 \ldots$)
and $n_f$ is the number of quark flavours with mass lower
than the renormalization scale $\mu$.  

Contrary to what happens in momentum-subtraction schemes ($MO$)
this beta function and the quark mass anomalous dimension 
are quark mass independent.
Thus, the Appelquist-Carazzone theorem~\cite{appelquist.carazzone:75},
that states that eventually the heavy particles
decouple at each order in perturbation theory, is not
realized in a trivial way since
coupling constants, beta functions and quark-mass anomalous 
dimensions do not exhibit it.
To obtain decoupling in the \msbar scheme we need to build
in the decoupling region, $\mu \ll m$, with $m$ the mass 
of the heavy particle, an effective field theory
~\cite{weinberg:80}
that behaves as if only the light degrees of freedom were present.
Matching conditions connect the parameters of the renormalized 
low-energy effective Lagrangian with the parameters of the full theory.
Power suppressed corrections of order $1/m$ 
contribute to physical observables only through higher order
operators but do not affect the matching conditions for the
coupling constant and quark masses.
The decoupling of the heavy particles
is fulfilled in physical quantities once they are expressed in
terms of the couplings in the effective theory.

Some time ago it was checked explicitly at
three loops~\cite{rodrigo.santamaria:93} that, when the appropriate 
matching conditions are taken into account, the evolution of the strong 
coupling constant from low energies to high energies does not depend on
the particular choice of the energy scale used to
pass a heavy quark threshold. The residual dependences that appear in
the perturbative calculation are just an estimate of the effects of the
higher order corrections.
Very recently~\cite{chetyrkin.kniehl.ea:97}
the analysis has been extended to four loops and the appropriate coefficients
have been computed. 

In this paper we use the recently calculated four-loop 
\msbar scheme QCD beta function and the quark mass anomalous dimension
\cite{ritbergen.vermaseren.ea:97,chetyrkin:97,vermaseren.larin.ea:97}
to obtain the logarithmic pieces in the matching conditions from a different
point of view.
Then, we obtain a very convenient analytic form for
the running of the QCD coupling constant at four loops and compare it with
other solutions. Finally, we use these results and the non-logarithmic
coefficients computed in~\cite{chetyrkin.kniehl.ea:97} to analyze the
impact of the matching conditions on the error induced in $\alpha_s(m_Z)$
if it is obtained from $\as(m_\tau)$ when passing the thresholds of the 
$c$ and the $b$ quarks. 

Matching conditions in QCD relate 
the strong coupling constant, $a_{n_f}$,
and the running mass of the light quarks, $\bar{m}_{q,n_f}$,
in the full theory with $n_f$ flavours
with the effective strong coupling constant, $a_{n_f-1}$,
and the effective light quark masses, $\bar{m}_{q,n_f-1}$,
of the effective theory with $n_f-1$ flavours 
through a power series in $a_{n_f-1}$
\bea
   a_{n_f}(\mu_{th}) &=& a_{n_f-1}(\mu_{th}) \left[ 1
+ \sum_{k=1}^{\infty} C_k(x) a_{n_f-1}^k(\mu_{th}) \right]~,
\label{matchalfa} \\
   \bar{m}_{q,n_f}(\mu_{th}) &=& \bar{m}_{q,n_f-1}(\mu_{th}) \left[ 1
+ \sum_{k=1}^{\infty} H_k(x) a_{n_f-1}^k(\mu_{th}) \right]~,
\label{matchmass}
\eea
with coefficients that depend on $x = \log(\mu^2_{th}/m^2)$
where $m$ is some RG-invariant mass of the heavy quark (for instance the 
RG-invariant-\msbar mass, $\bar{m}(\bar{m})$, or the perturbative pole mass 
$M$) that has been integrated out at the energy scale $\mu_{th}$.
In order to obtain a good approximation using only
the first few terms in the perturbative
expansion, we have to evaluate
matching conditions in a region where $\mu_{th}/m \sim O(1)$.
However, the result of these calculations should not depend
on exactly which $\mu_{th}$ is chosen.

Note that in contrast to other
analysis~\cite{chetyrkin.kniehl.ea:97,bernreuther:82,larin.ritbergen.ea:95}
where the effective couplings are expressed in terms of the
couplings of the full theory we directly write the inverted relation
since we are interested in the evolution of the QCD Lagrangian parameters 
from low energies to high energies.
Note also that in order to simplify as much as possible
the matching conditions we have taken as a reference mass, $m$,
a RG invariant mass instead of the running mass $\bar{m}(\mu_{th})$
evaluated at the threshold scale $\mu_{th}$. This makes matching conditions
for the $\alpha_s$'s independent of the anomalous dimensions.

The functions $C_k$ and $H_k$ are, in general, polynomials in $x$.
The coefficients multiplying the logarithms of 
the heavy quark mass are determined just by the RG,
that is, they are a function of the beta function and 
the quark mass anomalous dimension of both the effective theory
with $n_f-1$ flavours and the full theory with $n_f$ flavours.
The non-logarithmic coefficients, however, have to be be evaluated
explicitly for each particular renormalization scheme.

We apply the renormalization group equations, \eq{RGE}, 
to both sides of \eq{matchalfa} and \eq{matchmass}.
Identifying order by order in the effective strong coupling
constant, $a_{n_f-1}$,
we obtain for the $C_k$ and the $H_k$ functions
a set of coupled first-order linear differential equations
depending only on the beta and the gamma
functions of the full and the effective
theories. By solving them, and using the
known beta functions
we find for the $C_k$ functions
\bea
C_1 &=& \frac{x}{6}, \qquad
C_2 = c_{2,0}+\frac{19}{24}x+\frac{x^2}{36}~, \non \\
C_3 &=& c_{3,0} + \left(\frac{241}{54}+\frac{13}{4}c_{2,0} 
- \left( \frac{325}{1728} + \frac{c_{2,0}}{6} \right) n_f \right) x 
+ \frac{511}{576} x^2+ \frac{x^3}{216}~, 
\eea
while for the $H_k$ functions we obtain
\bea
H_1 &=& 0, \qquad
H_2 =  d_{2,0}+\frac{5}{36}x-\frac{x^2}{12}~, \non \\
H_3 &=& d_{3,0} + \left( \frac{1627}{1296}-c_{2,0}+\frac{35}{6} d_{2,0}
+\left( \frac{35}{648}-\frac{d_{2,0}}{3} \right) n_f
+\frac{5}{6}\zeta_3 \right) x \non \\
&-& \frac{299}{432} x^2  
- \left( \frac{37}{216} - \frac{n_f}{108} \right) x^3~, 
\eea
where $c_{2,0}$, $c_{3,0}$, $d_{2,0}$ and $d_{3,0}$ are
arbitrary constants coming from the integration of the differential equations.
They depend on the renormalization scheme and on the RG-invariant
reference mass $m$ chosen. They can be determined only
by evaluating some Green functions with both the full and the effective
theories, in a particular mass-independent renormalization scheme, and then
require they are the same, up to terms $O(1/m)$,
for values of the renormalization scale just around the threshold.
Note that, in order to simplify the results, we have set $c_{1,0}=d_{1,0}=0$,
which is the \msbar result with the usual dimensional regularization
prescription, $\mathrm{Tr}\{I\}=4$, with
the trace taken in Dirac space.

If the RG-invariant-\msbar mass is used as a reference scale, that is
$m=\bar{m}(\bar{m})$ the coefficients one obtains are
\cite{chetyrkin.kniehl.ea:97,larin.ritbergen.ea:95}
\begin{equation}
c_{2,0}=-\frac{11}{72}\ ,\ \ \ 
c_{3,0}=\frac{82043}{27648}\zeta_3-\frac{575263}{124416}+
\frac{2633}{31104} n_f\ ,\ \ \ 
d_{2,0}=-\frac{89}{432}~.  
\label{eq:cs-msbar}
\end{equation}

If the pole mass is used as a reference scale, that is
$m=M$ the coefficients one obtains are
\cite{chetyrkin.kniehl.ea:97}
\begin{eqnarray}
c_{2,0} &=&\frac{7}{24}\non\\ 
c_{3,0} &=& \frac{80507}{27648}\zeta_3+
\frac{1}{9}\zeta_2\left(2\log(2)+7\right)+\frac{68849}{124416}
-\frac{n_f}{9}\left(\zeta_2+\frac{2479}{3456}\right)~. 
\label{eq:cs-mpole}
\end{eqnarray}
while $d_{2,0}$ does not change, and $d_{3,0}$, when known, has to 
be shifted by a factor $+10/27$.

Using these coefficients we find complete agreement with
ref.~\cite{chetyrkin.kniehl.ea:97} also for the logarithmic contributions
to the matching for both elections of the reference scale. Note that
the authors of ref.~\cite{chetyrkin.kniehl.ea:97}  present the inverse 
relations.

For consistency, matching conditions at $n$ loops have to be 
considered together with running of the 
\msbar parameters at $n+1$ loops.
The $C_3$ and the $H_3$ coefficients depend on at most the 
three loop beta function and the three loop quark mass
anomalous dimension.
However, they have to be used together with running at four 
loops, where the recently calculated four loop beta 
function and the quark mass anomalous dimension enter the game.

Knowing the beta and gamma functions at four loops we could also obtain
the logarithmic pieces of $C_4$ and $H_4$, however since the coefficients
$c_{4,0}$ and $d_{4,0}$ are not known in any particular scheme 
and since these matching conditions should be used together with running at
five loops, which is also unknown, we do not present them.

In the following we sketch the formulae used to compute the running
of $\alpha_s$ at four loops, which is the required order if matching
conditions are used at three loops.

We solve the full four-loop RG equation for the strong
coupling constant as an expansion in the
solution of the two-loop RG equation. In \cite{rodrigo.santamaria:93}
we obtained an expression for the the running of the strong coupling
constant as an expansion of the strong coupling constant obtained
at one loop. We now improve that expression by resumming some of
the leading dependences proportional to $\beta_1$. This amounts to expanding
around the approximate two-loop solution instead of the one-loop solution. 

At the required order we have
\beq
   a(\mu) = a^{(2)}(\mu) \left( 1
                + c_2(\mu) (a^{(2)}(\mu))^2
                + c_3(\mu) (a^{(2)}(\mu))^3 \right)~,
\label{alfarun} 
\eeq
where $a^{(2)}(\mu)$ is the approximate two-loop 
solution
\beq
   a^{(2)}(\mu) = \frac{a(\mu_0)}
   {K+b_1 a(\mu_0) L +b_1^2 a(\mu_0)^2 (1-K+L )/K}
\eeq
and
\bea
c_2(\mu) &=&  b_2 (1-K)~,\non \\ 
c_3(\mu) &=& \frac{b_3}{2} \left(1-K^2\right) 
  + b_1 b_2 K \left(K-1-L\right)
 + \frac{b_1^3}{2} \left(L^2-(1-K)^2\right)~, 
\eea
with $K =1+\beta_0 \: a(\mu_0) \: \log(\mu^2/\mu_0^2)$, $L=\log K$ and
$b_k = \beta_k/\beta_0$.

Although these expressions are slightly more complicated than the usual
expansion in $1/\log(\Lambda_{QCD}) $\cite{chetyrkin.kniehl.ea:97}, they
are more convenient because the coupling constant at an arbitrary scale
is given explicitly in terms of the strong coupling constant at some 
reference scale $\mu_0$, which
usually one takes equal to $m_Z$ or to $m_\tau$. The standard formulae, 
however,
\mafigura{8 cm}{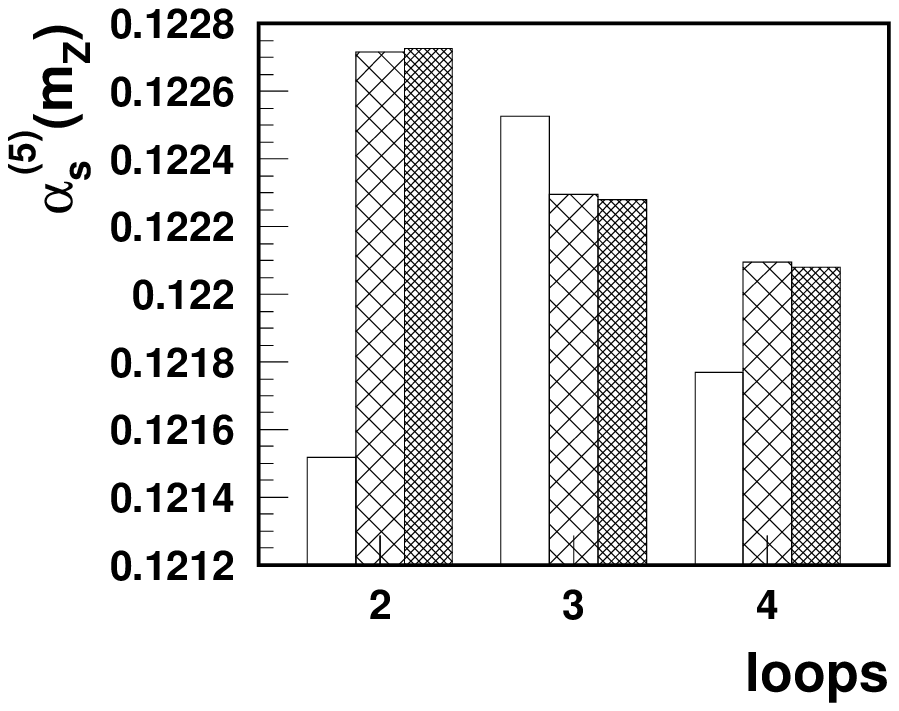}{Values one obtains for $\alpha^{(5)}_s(m_Z)$
starting from $\alpha^{(5)}_s(m_\tau)=0.336$ at 2,3 and 4 loops by using the
different approximations discussed in the text:
i) the usual $1/\log(\Lambda_{QCD})$
expansion \cite{chetyrkin.kniehl.ea:97} (white),
ii) numerical integration of the
renormalization group equation (soft hatching),
iii) our expansion in \eq{alfarun} (hard hatching).
}{fig:running}
are expressed in terms of $\Lambda_{QCD}$ which has to be adjusted to
the initial conditions for $\alpha_s$. So, the value of $\Lambda_{QCD}$
to be used depends on the number of light flavours considered
and on the number of loops considered. Moreover it seems that
the $1/\log(\Lambda_{QCD})$ expansion converges more slowly than
our expansion. We show in fig.~\ref{fig:running} the values
obtained for $\alpha^{(5)}_s(m_Z)$ starting from $\alpha^{(5)}_s(m_\tau)=0.336$
by using three different solutions of the RG at 2,3 and 4 loops : 
i) the usual $1/\log(\Lambda_{QCD})$
expansion \cite{chetyrkin.kniehl.ea:97} (white),
ii) numerical integration of the
renormalization group equation (soft hatching),
iii) our expansion in \eq{alfarun} (hard hatching).
Clearly our expansion gives much closer results to the numerical
integration and seems to converge faster than the $1/\log(\Lambda_{QCD})$
expansion. Therefore we will use it in the following.

Let us turn to the phenomenological applications of the matching
conditions and the four-loop running solutions. 

The most precise determinations of $\alpha_s$ are obtained from
hadronic $\tau$ decays~\cite{pich:97} at rather low energies  and
from hadronic $Z$ decays at LEP energies~\cite{lep} . To
compare these two results one has to connect the strong coupling constant
in a theory with three flavours at a scale
$\mu=m_\tau$  with the strong coupling constant in a theory with
five flavours at a scale $\mu=m_Z$. Therefore, two thresholds have to
be passed, the threshold of the $c$-quark and the threshold of the
$b$-quark. Moreover, running between a wide range of scales has to be
performed. Note that, although $m_c < m_\tau$,
results for $\alpha_s(m_\tau)$ are usually presented in a theory with
only three quark flavours. That is because $c$-quarks cannot be
really produced in $\tau$ decays and they only enter in loops. Therefore
it is appropriate to use an effective theory in which the $c$-quark
has been integrated out. Power corrections of the form $m^2_\tau/m^2_c$
can be included in the effective theory and have been
computed \cite{chetyrkin:93,larin.ritbergen.ea:95}. They are very
small and they are taken into account in the extracted value of
$\as(m_\tau)$.

Given the present accuracy of both experimental measurements and
theoretical calculations of hadronic decays of the $\tau$ lepton, 
it is important to calculate very precisely the connection between 
coupling constants when passing thresholds and to estimate the remaining 
errors in the calculation. It is in this analysis where the three 
loop-matching conditions for $\alpha_s$ and four-loop running 
studied above are relevant. 

In the following we obtain $\as(m_Z)$ by using as starting point
$\as(m_\tau)$ and we estimate the residual errors due to the matching
conditions and running for the different approximations used.
We follow the same procedure as in~\cite{rodrigo.santamaria:93}.
At low energies, $\mu=m_\tau=1777.0\pm 0.3~MeV$ we know \cite{pich:97}
$\as(m_\tau) \equiv \as^{(3)}(m_\tau)  = 0.35\pm 0.02$.
From this we can obtain $\as^{(3)}(\mu_{th}^c)$ at some matching point
$\mu_{th}^c$ around $\bar{m}_c$ by using the renormalization group with
$n_f=3$, then, by using \eq{matchalfa} with $m=\bar{m}_c$ and $n_f=4$ we
obtain $\as^{(4)}(\mu_{th}^c)$. Now we use again the renormalization group
with $n_f=4$ to obtain $\as^{(4)}(\mu_{th}^b)$ at some matching point
$\mu_{th}^b$ around $\bar{m}_b$ and use again \eq{matchalfa} with
$m=\bar{m}_b$ and $n_f=5$ to obtain $\as^{(5)}(\mu_{th}^b)$. Finally
we use the renormalization group with $n_f=5$ to obtain
$\as^{(5)}(m_Z)\equiv \as(m_Z)$. The final result will depend on the
precise values used for $\mu_{th}^c$ and $\mu_{th}^b$ and this dependence
gives an estimate of the errors which arise because the truncation of the
perturbative series in the matching conditions. In addition, matching
conditions also depend on the masses of the quarks, and,
although they are very well known, their actual value can affect
the final result for $\as(m_Z)$. The induced error due to the 
uncertainty in the quark masses is dominated by the one-loop
matching equation. Then, we can estimate this 
error as $(\Delta \alpha_s(m_Z))/\alpha_s(m_Z) \approx
\alpha_s(m_q)/(3\pi) (\Delta m_q)/m_q$.
We use always as a reference scale the RG-invariant-\msbar mass of the quarks,
and therefore the coefficients in \eq{eq:cs-msbar}. For the quark masses
we take the last values in the literature: for the $b$-quark mass
$\bar{m}_b(\bar{m}_b) = 4.13 \pm 0.06~GeV$ \cite{jamin.pich:97}.
For the $c$-quark mass we take $\bar{m}_c(\bar{m}_c)=1.31 \pm 0.06~GeV$,
\cite{narison:?} (see also \cite{rodrigo:95}).

We study the effect of varying the scale at which matching is performed
independently for the $c$ and the $b$ quarks. 

\mafigura{8 cm}{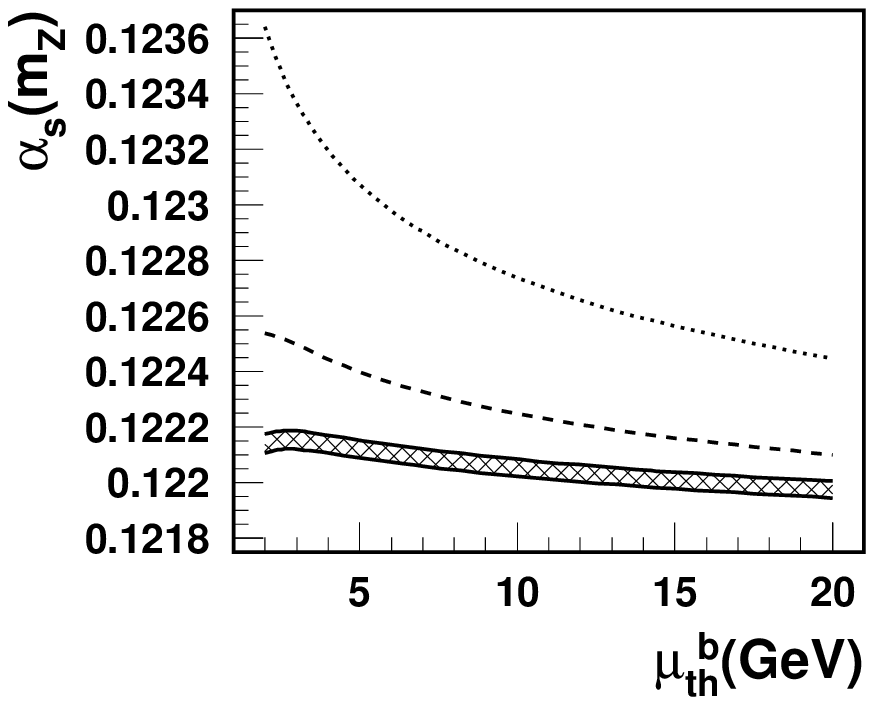}{
$\as(m_Z)$ obtained by running the coupling from $\as(m_\tau)=0.35$, 
as a function of the matching point taken to cross the $b$-quark threshold.
Dotted line: two-loop beta functions and one-loop matching conditions. 
Dashed line: three-loop beta functions and two-loop matching conditions. 
The hatched band is obtained with four-loop beta functions and three-loop
matching conditions when the $b$-quark mass is varied within its
error interval.}{fig:b}
First we fix $\mu_{th}^c=\bar{m}_c$ and vary $\mu_{th}^b$ in
$2-20$ GeV.
Figure~\ref{fig:b} shows (dotted line) our result for
two-loop running (only the first term of~\eq{alfarun} is
taken into account) and one-loop matching conditions (only the $C_1$
coefficient is considered). We plot with dashed line the
results for three-loop running and two-loop matching conditions
(two terms in~\eq{alfarun} and $C_2$ included in~\eq{matchalfa}).
For these two lines we took central values for the $b$-quark mass.
Finally, the hatched area gives the results for four-loop running and
three-loop matching conditions when the $b$-quark mass is varied within
its error interval. For the central value of the strong coupling 
constant extracted from tau decays we find that
varying the $b$-quark threshold scale in the range
$\mu_{th}^b = 2-20~GeV$ two-loop running and one-loop
matching conditions induce an error of $0.0006$ on the
strong coupling constant at the $Z$-boson mass scale.
With three-loop running and two-loop matching
conditions the error decreases to $0.0002$.
For four-loop running and three-loop matching
conditions we get and error for the central value of the
$b$-quark mass of $0.00009$. The uncertainty in the $b$-quark mass
induces in this case an additional error of $0.00003$.

To study the errors induced in passing the $c$-quark threshold
we fix $\mu_{th}^b=\bar{m}_b$ and vary $\mu_{th}^c$
in the range $\mu_{th}^c = 1-4 GeV$. Then, we find 
an induced error of $0.0005$, $0.0002$ and $0.0001$
to each order respectively. The uncertainty in the $c$-quark mass
introduces an additional error of $0.0001$.

To analyze the combined effect of passing the two thresholds we have
represented the  value of $\alpha_s(m_Z)$ obtained at four loops
as a function of $\mu^c_{th}$ and $\mu^b_{th}$ as a contour plot in
fig.~\ref{fig:contour} by taking $\log (\mu^b_{th}/1GeV)$ and 
$\log (\mu^c_{th}/1GeV)$ in the
$x$ and $y$ axis respectively. The different contour lines are obtained for
$\as(m_Z)= 0.12214$ to $\as(m_Z)= 0.12182$ in steps of $0.00004$.
\mafigura{8 cm}{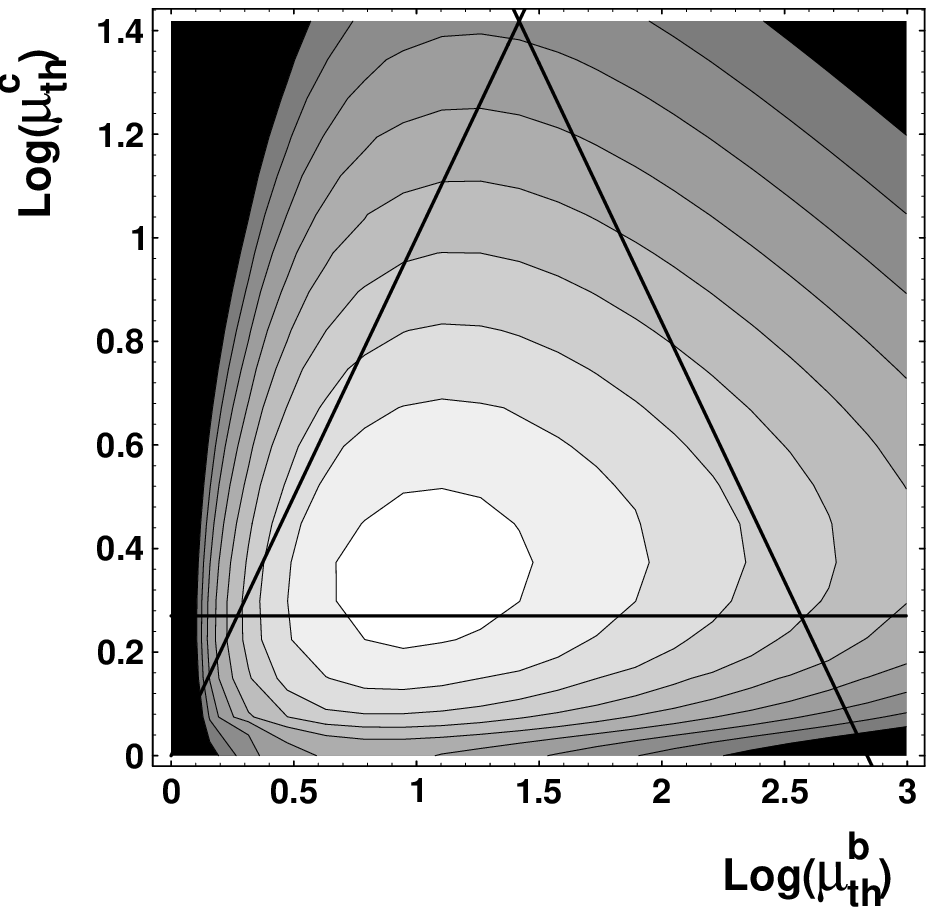}{
$\alpha_s(m_Z)$ as a function of the matching point scales $\mu^c_{th}$
and $\mu^b_{th}$. The contours are for $\as(m_Z)= 0.12214$ to
$\as(m_Z)= 0.12182$ in steps of $0.00004$.
The area within the strait lines is obtained by taking
$\bar{m}_c < \mu^c_{th} < \mu^b_{th} < \bar{m}_b^2/\mu^c_{th}$.
}{fig:contour}
We see that there is a maximum approximately for
$\alpha_s(m_Z)=0.1222$. The minimum depends on how far away from
$\bar{m}_b$ and $\bar{m}_c$ we take the matching scales $\mu^b_{th}$ and
$\mu^c_{th}$. We choose $\bar{m}_c < \mu^c_{th} < \mu^b_{th} < \bar{m}_b^2/\mu^c_{th}$,
which are represented by the three strait lines in fig.~\ref{fig:contour}.
Then we obtain an estimate of the error due to the unknowledge of the
scales $\mu^c_{th}$ and $\mu^b_{th}$ of about $\Delta \alpha_s(m_Z) =0.0002$ 
by taking half the difference between the maximum and the minimum values of
$\alpha_s(m_Z)$. 
If in addition to this
we also consider the error induced because the uncertainty in the input
values of the masses
of the $c$ and the $b$ quarks, we obtain that the total error
in $\alpha_s(m_Z)$ due to the matching procedure  is
$\Delta \alpha_s(m_Z) = 0.0003$. Finally we can also include the error
in the extracted value of $\alpha_s(m_\tau)$, which
is the dominant one. We obtain the values of
$\alpha_s(m_Z)$ starting from $\alpha_s(m_\tau) = 0.35 \pm 0.02$
and take the average of the maximum and minimum to get the central value 
and half of the difference to obtain the error. This gives 
$\Delta \alpha_s(m_Z) = 0.0021$. Combining lineally all errors we obtain
\beq
\alpha_s(m_Z) = 0.1219 \pm 0.0024
\eeq
Clearly, considering matching conditions at three loops and running at
four loops reduces the errors due to the matching procedure by a 
factor two and renders them below 0.0005 in $\as(m_Z)$. 
This has to be compared with the full error which is of about $0.0025$.

By using the recent four-loop QCD calculations we have obtained the
matching conditions for $\as(\mu)$ and $\bar{m}_q(\mu)$ when crossing
a quark threshold. Then, we have solved the renormalization group
equations at four-loops and have obtained a very convenient analytic form for
the running of the QCD coupling constant. We have compared
the results obtained by using this solution with the results obtained 
by using the usual 
$1/\log(\Lambda_{QCD})$
expansion and with the numerical solution at the different orders and
found that it gives a good approximation to the numerical solution and
converges faster than the usual $1/\log(\Lambda_{QCD})$ expansion.
Finally we have used these results to study the
effect of the $c$- and the $b$-quark
thresholds in the evolution of the strong
coupling constant from $\mu=m_\tau$ to $\mu=m_Z$.
This analysis is very important given the present accuracy of
both determinations of the QCD coupling constant, $\alpha_s(m_\tau)$
and $\alpha_s(m_Z)$. We found that the total error induced
in $\as(m_Z)$ starting from $\as(m_\tau)$ due to the matching
and running procedures is $0.0003$ when matching and running
evolution are used at four loops.

\vspace{1cm}

Work supported in part by CICYT (Spain) under grant AEN-96-1718 and
in part by DFG Projekt No. Ku 502/8-1 (Germany). 
A.S. would like to thank SISSA, Trieste (Italy) for the warm hospitality 
during his visit there, where part of this work has been done.
The work of G.R. has also been supported in part by CSIC-Fundaci\'o Bancaixa.

\end{document}